\documentclass[a4paper,11pt]{article}
\usepackage{pos}
\usepackage[utf8]{inputenc}
\usepackage[T1]{fontenc}
\usepackage{physics}
\usepackage{amsfonts}
\usepackage{caption}
\usepackage{subcaption}
\usepackage{array}
\usepackage{indentfirst} 
\usepackage{amsmath}
\usepackage[toc,page]{appendix}
\usepackage{mathtools}
\usepackage{hyperref}
\usepackage{verbatim}
\usepackage{graphicx} 
\usepackage{booktabs} 
\usepackage{xcolor}
\usepackage{soul}
\usepackage{bbding}
\usepackage{physics}

\newcommand{\Lagr}{\mathcal{L}}

\title{Curvature-induced electroweak symmetry breaking and phase transition of a Higgs-portal dark scalar field}
\ShortTitle{Curvature-induced EWSB and PT of a Higgs-portal dark scalar field}

\author*[a,b,c,d]{Andreas Mantziris}

\affiliation[a]{Departamento de Física e Astronomia, Faculdade de Ciências, Universidade do Porto, Rua do Campo Alegre s/n, 4169-007 Porto, Portugal}

\affiliation[b]{Centro de F\'isica das Universidades do Minho e do Porto, LaPMET, Universidade do Porto, Rua do Campo Alegre s/n, 4169-007 Porto, Portugal}

\affiliation[c]{Univ Coimbra, Faculdade de Ci\^encias e Tecnologia da Universidade de Coimbra, Rua Larga, 3004-516 Coimbra, Portugal}

\affiliation[d]{Physics Department, University of Ioannina, 45110, Ioannina, Greece}
  
\emailAdd{andreas.mantziris@uc.pt}

\abstract{This overview of the study \cite{Mantziris:2024uzz}, regarding the possibility of generating gravitational waves from a curvature-induced phase transition of a non-minimally coupled scalar dark matter field with a Higgs-portal, was showcased at the ``Workshop on Standard Model and Beyond 2025'' of the Corfu Summer Institute 2025. The phase transition dynamics during the transition from inflation to kination were calculated for various inflationary scales, considering both positive and negative values of the non-minimal coupling, while also examining the potential for triggering electroweak symmetry breaking. Notably, kination enhances the GW amplitudes, significantly restricting the viable parameter space. While the GW spectra follow the usual rule for high-frequencies from high inflationary scales, certain regions of the parameter space allow for a potential detection with future experiments.}

\FullConference{Proceedings of the Corfu Summer Institute 2025 "School and Workshops on Elementary Particle Physics and Gravity" (CORFU2025)\\
27 April - 28 September, 2025\\
Corfu, Greece\\
}

\begin{document}
\maketitle

\section{Motivation and overview}

Understanding the laws governing the Universe is the central pursuit of theoretical physics, with the Standard Model (SM) of particle physics standing tall as its cornerstone. Despite its remarkable success, the SM cannot explain several phenomena in particle physics and cosmology, suggesting the need for physics Beyond the SM (BSM) \cite{Bose:2022obr}. Numerous BSM frameworks \cite{Fox:2022tzz} attempt to address these shortcomings, while the limitations of current colliders motivate us to seek complementary routes for new insights, like early universe cosmology \cite{Eichhorn:2018cyr} and Gravitational Wave (GW) astronomy \cite{Caprini:2025trt}. A longstanding issue in cosmology, unexplained by the SM is the elusive component comprising $\sim26\%$ of the Universe dubbed Dark Matter (DM). DM appears to interact only gravitationally and is responsible for the large scale structure that ``seeds'' galaxy formation \cite{Bertone:2004pz}. Many BSM theories propose compelling DM candidates, which are often linked to early universe phenomena, like Phase Transitions (PTs) and symmetry-breaking processes \cite{Cirelli:2024ssz}. Such phenomena with characteristic observational signatures offer promising probes of the dynamics of the underlying mechanisms through the study of both the microscopic physics of particle interactions and the macroscopic imprints on cosmological observables.

Recent advances in GW physics present a unique opportunity to probe the early universe and the exotic particles that it can accommodate via distinctive signals from inflation, PTs, and topological defects \cite{LISACosmologyWorkingGroup:2022jok}. Since GWs travel unhindered after emission, they offer a direct window into the primordial universe, shedding light on the dynamics of quantum fields in high-energy settings. In particular, cosmological PTs are of significant interest for two main reasons: first, they are expectedly prevalent due to the Electroweak (EW) and Quantum Chromodynamics crossover transitions; second, many BSM theories predict PTs that can generate detectable GW spectra with current and upcoming surveys \cite{Caprini:2018mtu}, such as LIGO \cite{LIGOScientific:2019vic}, LISA \cite{LISA:2024hlh}, AEDGE \cite{AEDGE:2019nxb} and the Einstein Telescope (ET) \cite{Punturo:2010zz}. These observations will provide a significant advancement in our understanding of our fundamental theories by highlighting the proper avenues beyond the SM, while also deciphering the behaviour of the Universe at its very early moments. 

The main paradigm for the description of the primordial universe is inflation, a period of accelerated expansion with an exponential growth of the scale factor resulting from a scalar field's approximately constant potential energy \cite{Starobinsky:1980te, Guth:1980zm, Sato:1980yn}. At the end of inflation, the inflaton's energy is effectively transferred into SM particles via a thermal process that (re)heats the Universe into the Hot Big Bang (HBB) epoch, where the details of this (p)reheating period being elusive \cite{Barman:2025lvk}. While inflation solves some of the long-standing problems of the HBB model (horizon, flatness, and relic abundance), it provides the mechanism for the initial quantum fluctuations in the matter density field to grow and seed the large-scale structure of the late universe \cite{Liddle:2000cg}, as evidenced by the Cosmic Microwave Background (CMB) anisotropies \cite{Planck:2019evm}.


This work provides a pedagogical review of Ref.~\cite{Mantziris:2024uzz}, where these considerations motivated us to investigate a minimal renormalizable BSM DM extension \cite{Cosme:2018wfh}, in the form of a non-minimally coupled Higgs-portal scalar field defined in Sec. \ref{sec:BSM}, and its potential gravitational radiation from a PT. The curvature-induced PT of the dark sector under study takes place after inflation \cite{Kierkla:2023uzo} during the kination \cite{Joyce:1996cp} / deflation \cite{Spokoiny:1993kt} epoch, where the inflaton's kinetic energy dominates. Extended periods of kination are naturally found in scenarios of quintessential inflation, where the inflaton is responsible also for the late-time accelerated expansion of the Universe due to a second plateau in its potential \cite{Bettoni:2021qfs}. In contrast to the widely studied cases of thermal PTs (see for example Refs. \cite{Alanne:2020jwx, Ellis:2022lft, Koutroulis:2023wit, Caprini:2024hue, Krajewski:2024xuz, Barman:2026kab} and references therein), that occur at relatively low energy scales \cite{Linde:1980tt, Linde:1981zj} due to the decreasing temperature of the Universe after radiation-domination, we consider a non-thermal PT that can illuminate earlier moments and complement the insights from other observables without the interference of low-frequency sources \cite{Cutting:2018tjt, Bettoni:2019dcw, Bettoni:2021zhq, Laverda:2023uqv, Buckley:2024nen, Bettoni:2024ixe, Laverda:2025pmg, Rubio:2025egw, Laverda:2026slq}. Furthermore, this model can induce the Breaking of EW Symmetry (EWSB) via the evolution of spacetime curvature, assuming sufficiently low (re)heating temperatures. A potential detection of such GW signals could decipher the parameter space of the BSM model and pinpoint to the relevant inflationary scale, as discussed in Sec. \ref{sec:GWs}.

\section{Cosmological evolution of a dark scalar in the early universe} \label{sec:BSM}
\subsection{Spectator Higgs-portal field non-minimally coupled to gravity}

\begin{figure}[h]
\centering
\includegraphics[scale=0.9]{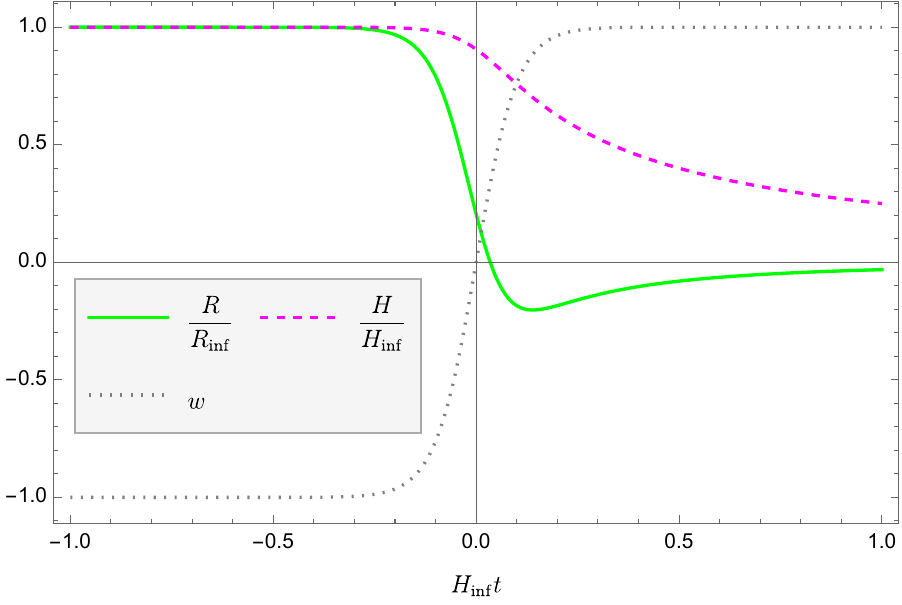}
\caption{Transition from inflation to kination for a quintessential-like model regarding the Ricci scalar, the Hubble rate, and the EoS parameter for a generic parametrisation $w(t)=\tanh{[\beta_w (t-t_0)]}$ with $\beta_w = 10 H_{\rm inf}$.}
\label{fig:inf}
\end{figure}

Following Ref. \cite{Cosme:2018wfh}, we extended the SM with a non-minimally coupled massless scalar field that acts as cold DM in the late universe. We modified the $\mathbb{Z}_2$-symmetric model of Ref. \cite{Cosme:2018wfh} by adding a renormalizable cubic term in the potential, so that it can readily exhibit a double-well shape and host strong first-order PTs via the mechanism proposed in Ref. \cite{Kierkla:2023uzo}. Hence, the action of our theory consists of Einstein gravity, the inflaton $\varphi$, the SM Higgs $h$ (we omit the rest of the SM for brevity), and the dark scalar $\phi$, reading
\begin{align}
    S=\int d^4 x \sqrt{-g} \left[\frac{M_P^2 -\xi_h h^2}{2} R-\frac{1}{2}\partial_\mu \phi \partial^\mu \phi- V_\phi - \Lagr_h - \Lagr_{\varphi} \right] \,,
    \label{eq:S}
\end{align}
where $M_P = 2.435 \times 10^{18}$ GeV is the reduced Planck mass, $R$ is the Ricci scalar, $V_{\phi}$ is the BSM potential, and $\Lagr_h$($\Lagr_{\varphi}$) is the Higgs (inflaton) Lagrangian. It is imperative to include the non-minimally coupled terms of the fields to curvature, since they are necessary for the renormalizability of the theory, radiatively generated and unavoidable due to their running couplings $\xi$ \cite{Birrell:1982ix, Mantziris:2020rzh}. However, we neglect the $\xi_{\varphi}\varphi^2R$-term, as we are not limited to a specific ``new'' inflation model, as long as it exhibits its standard behaviour and accommodate some period of kination, as depicted in Fig. \ref{fig:inf}. More specifically, the Hubble rate $H$ is approximately constant during inflation and decreases with time\footnote{This is the generic behaviour of the Hubble rate in ``new'' and quintessential inflation scenarios, until the onset of kination, with the analytical expression obtained by solving the first slow-roll parameter (see Refs. \cite{Mantziris:2024uzz, Kierkla:2023uzo}).}, as the kinetic energy of the inflaton increases, and the Equation of State (EoS) parameter, relating pressure and density $w=p/\rho$, changes sign from $w(t_{\rm inf})=-1$ to $w(t_{\rm kin})=1$. In these slow-roll models, the curvature scalar switches sign as inflation ends, 
\begin{align}
    R(t) =3 \left[ 1-3w(t) \right] H^2(t) \, .
    \label{eq:R}
\end{align}

To keep the model as minimal as possible, we limited the theory-space of the BSM scalar singlet to renormalizable interactions with no higher-order terms,
\begin{align}
      V_{\phi}(\phi, h, R) = \frac{1}{2} \left(\xi_{\phi} R + \frac{g^2}{2} h^2 \right) \phi^2-\frac{\sigma}{3}\phi^3+\frac{\lambda_{\phi}}{4}\phi^4 \, ,
      \label{eq:Vphi}
\end{align}
where $g$ is the Higgs-portal coupling, $\sigma$ is the dimensionful cubic coupling, and $\lambda_{\phi}$ is the dark self-coupling. Although there is no $\mathbb{Z}_2$ symmetry in the dark sector, we can neglect linear $\phi$-terms as they can be absorbed via a field redefinition \cite{Adams:1993zs}. Furthermore, $\phi-\varphi$ interactions are also omitted since they do not affect the relative location of the potential's minima and do not alter the PT dynamics. The Higgs-inflaton terms were also ignored, considering that during inflation, the Higgs field is stuck at the origin $h=0$ and the inflaton is subdominant by the time of EWSB \cite{Mantziris:2022fuu}. 

The BSM theory (\ref{eq:Vphi}) acts as DM in the late universe via the Higgs-portal as in Ref. \cite{Cosme:2018wfh}, albeit with different coupling strengths. In contrast to Ref. \cite{Cosme:2018wfh}, where EWSB results from the dark scalar ``rolling'' down its potential, EWSB is induced by the non-minimally coupled Higgs term, if temperature does not restore the symmetry, $T_{\rm reh} \leq 80$ GeV. The PT mechanism of Ref. \cite{Kierkla:2023uzo}, with the evolving potential barrier and vacua from the dynamic curvature (\ref{eq:R}), endows the model with a stochastic GW background, that can constrain its parameter space and enrich its phenomenology prospects. Finally, a quintessential-like inflationary scenario implies that no oscillations of $R$ complicate the PT process and that a long kination era amplifies the GWs \cite{Bettoni:2021qfs}. However, this is not a necessary assumption, since our treatment applies to any scenario where $R$ changes sign, with the differences in the corresponding GWs and the phenomenology of each model.

\subsection{Cosmological phase transitions in vacuum}
As inflation finishes and the Ricci scalar changes sign, the potential (\ref{eq:Vphi}) can develop a barrier between a true and a false vacuum state, resulting in a double-well shape. In order to simplify the analysis, we can rewrite the potential in the reduced dimensionless form \cite{Adams:1993zs} as 
    \begin{align}
    \widetilde{V}(\tilde{\phi}, t) &= \frac{1}{4}\tilde{\phi}^4 - \tilde{\phi}^3 +\frac{\delta(t)}{2}\tilde{\phi}^2 \, , 
    \label{eq:Vtilde} \\
    \tilde{\phi} &= \frac{3\lambda_{\phi}}{\sigma}\phi \, , 
    \label{eq:phi_tilde}\\
    \delta (t) &= \frac{9 \lambda_{\phi} \left[ \xi_{\phi} R(t) + \frac{g^2}{2} h^2 \right]}{\sigma^2} \, ,
    \label{eq:delta}
    \end{align} 
where the only coupling follows the evolution of $R(t)$. Any $4^{\rm th}$-order polynomial potential can be written as Eq. (\ref{eq:Vtilde}), with the appropriate field and coupling redefinitions (\ref{eq:phi_tilde})-(\ref{eq:delta}) \cite{Adams:1993zs}. This approach allows us to study the evolution of the potential both analytically and numerically, since $\delta$ encodes all the PT dynamics and dimensionful quantities. Specifically, $\delta$ quantifies the shape of the potential between the ``extremum'' configurations of a vanishing barrier or two degenerate vacua, $0\leq \delta \leq 2$. Its strength can take other values, but this range encodes the relevant parameter space for a PT assuming that the false vacuum is fixed at the origin, $\Tilde{\phi}_{\rm fv}=0$. It is helpful to insert Eq. (\ref{eq:R}) into Eq. (\ref{eq:delta}), so that the dependence of the potential on the EoS parameter is made explicit, resulting in
\begin{align}
    \delta &= C \left[ \frac{1 - 3 w(t)}{2} \right] \left( \frac{H(t)}{H_{\rm inf}}\right)^2 \, , \\
    C &= 54 \lambda_{\phi} \xi_{\phi} \left( \frac{H_{\rm inf}}{\sigma} \right)^2 \, .
    \label{eq:C}
\end{align}

\begin{figure}[ht!]
\begin{minipage}{.5\linewidth}
\centering
\subfloat{\includegraphics[scale=.55]{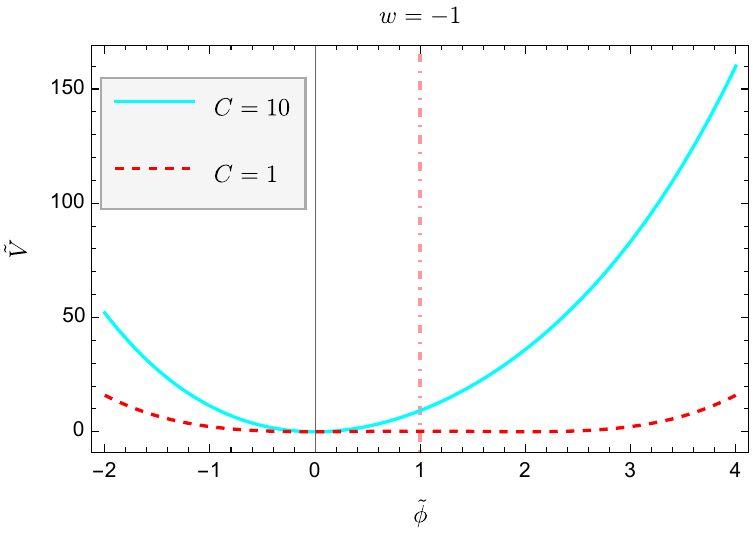}}
\end{minipage}
\begin{minipage}{.5\linewidth}
\centering
\subfloat{\includegraphics[scale=.55]{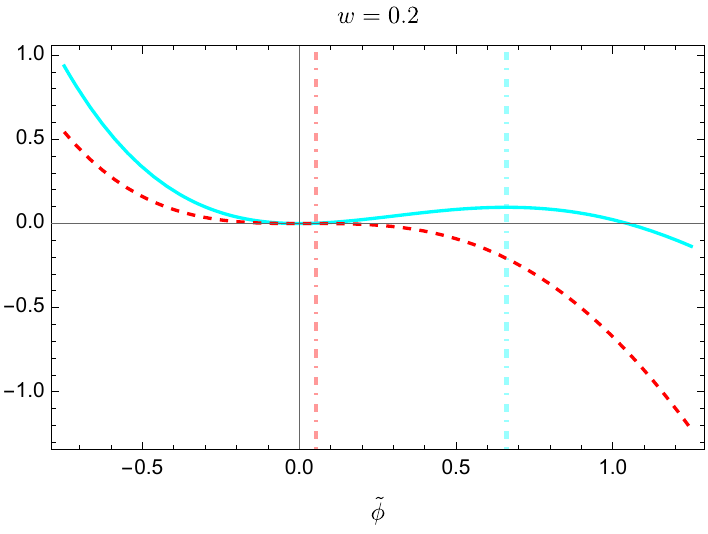}}
\end{minipage} \par\medskip
\begin{minipage}{.5\linewidth}
\centering
\subfloat{\includegraphics[scale=.55]{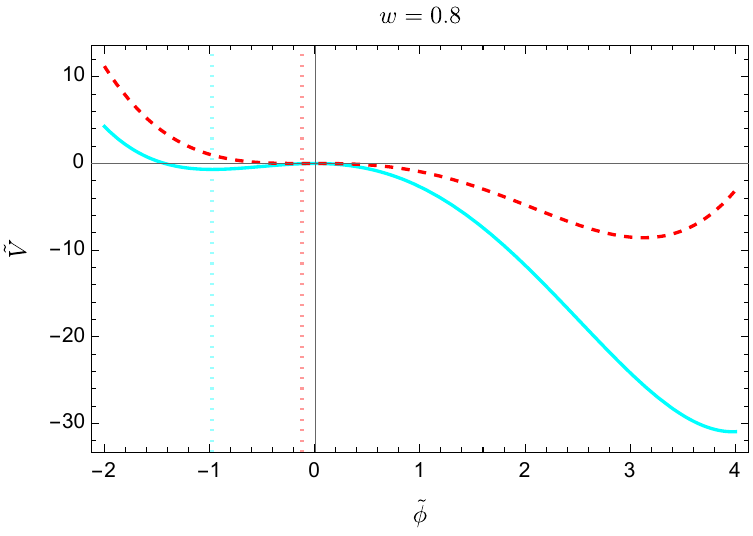}}
\end{minipage}
\begin{minipage}{.5\linewidth}
\centering
\subfloat{\includegraphics[scale=.55]{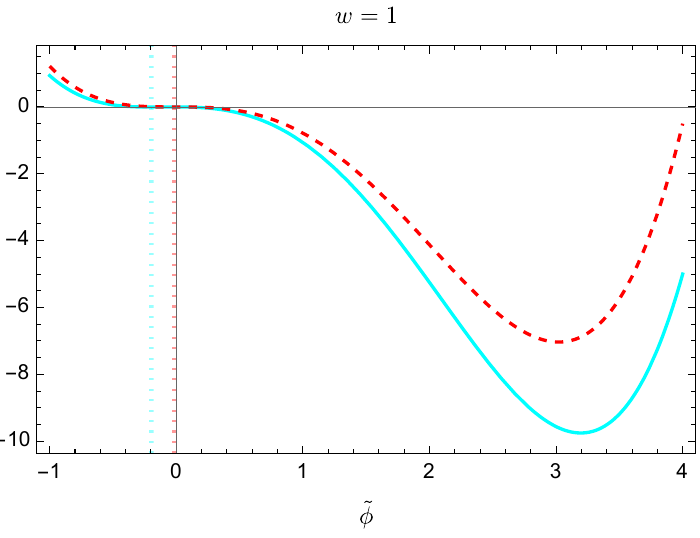}}
\end{minipage}
\caption{Evolution of the dark scalar potential for $\xi_{\phi}>0$ from inflation (upper left) to kination (lower right). The vertical dash-dotted and dotted lines correspond to the respective barriers and false vacua.}
\label{fig:V-xi-positive}
\end{figure}

\begin{figure}[ht!]
\begin{minipage}{.5\linewidth}
\centering
\subfloat{\label{fig:Vinf1}\includegraphics[scale=.55]{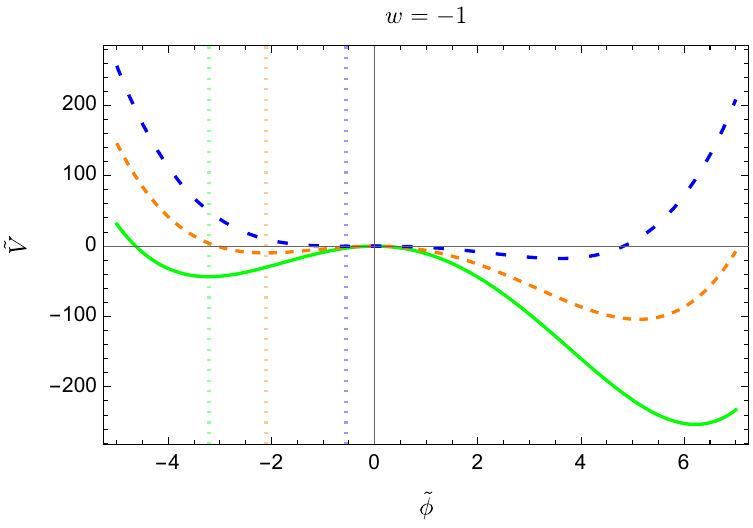}}
\end{minipage}
\begin{minipage}{.5\linewidth}
\centering
\subfloat{\label{fig:Vinf2}\includegraphics[scale=.55]{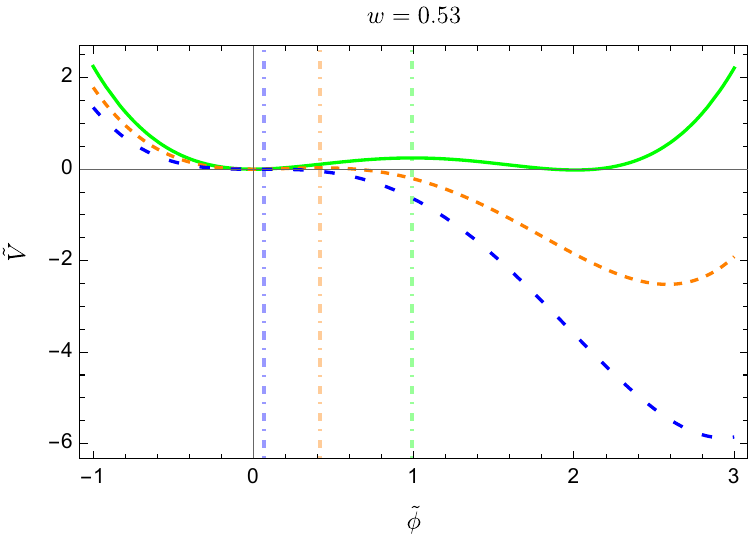}}
\end{minipage} \par\medskip
\begin{minipage}{.5\linewidth}
\centering
\subfloat{\label{fig:Vinf3}\includegraphics[scale=.55]{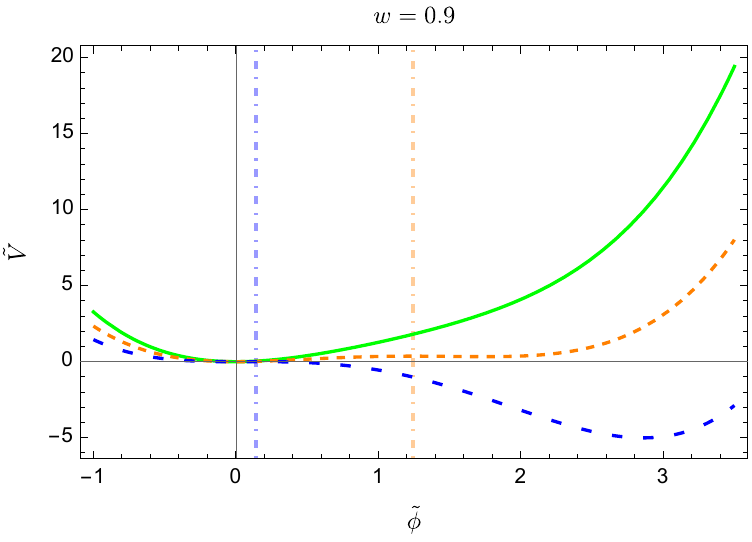}}
\end{minipage}
\begin{minipage}{.5\linewidth}
\centering
\subfloat{\label{fig:Vinf4}\includegraphics[scale=.55]{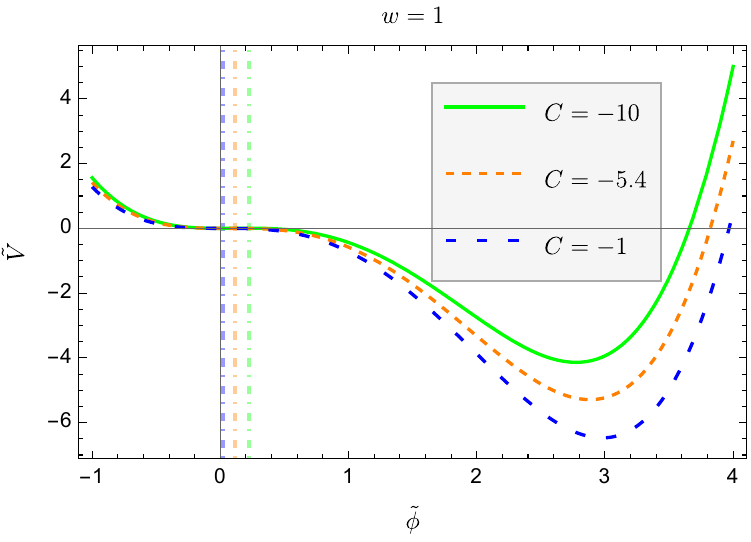}}
\end{minipage}
\caption{Evolution of the BSM potential for $\xi_{\phi}<0$ from inflation (upper left) to kination (lower right). The vertical dotted and dash-dotted lines correspond to the respective false vacua and barriers.}
\label{fig:V-xi-negative}
\end{figure}

The dark scalar dynamics for positive and negative values of the non-minimal coupling in this formalism are depicted in Figs. \ref{fig:V-xi-positive} and \ref{fig:V-xi-negative}. For $\xi_{\phi}>0$, the potential can feature no barrier or a true vacuum during inflation (top left plot of Fig. \ref{fig:V-xi-positive}). As the kinetic energy of the inflaton takes over its potential, a true vacuum develops at large field values with a barrier of decreasing height and width (top right panel of Fig. \ref{fig:V-xi-positive}). At some point, the barrier becomes thin enough so that a first-order PT takes place, and the field tunnels through the barrier from false to true vacuum. This leads to a runaway process in the surroundings, where $\phi$ follows the trajectory to the true vacuum, resulting in the formation of a spherically symmetric bubble that expands with approximately the speed of light, in the absence of radiation pressure \cite{Callan:1977pt, Coleman:1980aw}. In the late universe, the field settles in the true vacuum with a non-zero vacuum expectation value (vev) (bottom right plot of Fig. \ref{fig:V-xi-positive}).

In the case of $\xi_{\phi}<0$, the evolution of the potential is reversed. During inflation, the field lies at the high-field vev (top left panel of Fig. \ref{fig:V-xi-negative}). As time progresses, the true vacuum becomes shallower and the symmetry may be restored (top right and bottom left plots of Fig. \ref{fig:V-xi-negative}), depending on the couplings in Eq. (\ref{eq:C}). Within a viable patch of the parameter space, e.g. for the reference point $C=-5.4$, the field tunnels through the barrier towards the origin\footnote{This PT trajectory is directed in the opposite manner from the usual prescription, where a redefinition of the potential is necessary to use the framework of Ref. \cite{Adams:1993zs}, detailed in Appendix B of Ref. \cite{Mantziris:2024uzz}. \label{foot:2}}, if the PT proceeds fast enough compared to the background's expansion. In the late universe, the potential regains its broken shape with no barrier to separate the true vacuum from the origin, and the scalar rolling towards its high-field vev (bottom right panel of Fig.\ref{fig:V-xi-negative}). The late behaviour of the BSM field resembles a second-order transition, with no GW production, but nonetheless ensures that the dark scalar can acquire mass and play the role of DM after EWSB, as we will see in the next section.

To determine the susceptibility of a scalar field to undergo a PT, we need to compare the expansion of the background spacetime $H$ with the (vacuum) decay rate,
\begin{align}
    \Gamma = m^4 \left(\frac{S_4^E}{2\pi}\right)^2 e^{-S_4^E -\frac{1}{2} \Sigma_4} \, ,
\end{align}
where $m^2$ is the effective quadratic coupling of the scalar, and the regularised sum over multipoles,
\begin{align}
    \Sigma_4 (\delta) = \frac{53.9926 - 47.6801 \delta +11.0134 \delta^2 +0.3358 \delta^3 + 0.4197 \delta^4 - 0.2938 \delta^5}{(2 - \delta)^3} \, ,
\end{align}
is the 1-loop contribution to the $O(4)$ Euclidean action with $\alpha_{1,2,3}$ being numerical factors \cite{Adams:1993zs, Matteini:2024xvg},
\begin{align}
    S_4^E (\delta) = \frac{4 \pi^2}{3 \lambda_{\phi}}(2-\delta)^{-3}\left[\alpha_1\delta + \alpha_2 \delta^2 + \alpha_3 \delta^3\right] \, .
\end{align} 
 Depending on the sign of $\xi_{\phi}$, and thus the trajectory to the true vacuum (footnote \ref{foot:2}), we have 
\begin{align}
    m^2_{\xi_{\phi}>0} (R, h=0) &= \xi_{\phi} R \, , \\
    m^2_{\xi_{\phi}<0} (R, h=0) &= \frac{\sigma^2}{4 \lambda_{\phi}} \left(1 + \sqrt{1 + \frac{4 \lambda_{\phi} |\xi_{\phi}| R}{\sigma^2}}\right) + |\xi_{\phi}| R \, ,
\end{align}
with the corresponding quadratic coupling in the reduced dimensionless form for $\xi_{\phi}<0$ being
\begin{align}
\tilde{\delta}(R, h=0) = \left( \frac{8 \lambda_{\phi}}{\sigma^2 } \right) \frac{ \frac{\sigma^2}{4 \lambda_{\phi}} \left(1 + \sqrt{1 + \frac{4 \lambda_{\phi} |\xi_{\phi}| R}{\sigma^2}}\right) + |\xi_{\phi}| R }{\left( \frac{1}{3} + \sqrt{1 + \frac{4 \lambda_{\phi} |\xi_{\phi}| R}{\sigma^2}}\right)^2} = \frac{2 \left( 1 - \frac{4 \delta}{9} + \sqrt{1 - \frac{4 \delta}{9}}\right)}{\left( \frac{1}{3} + \sqrt{1 - \frac{4 \delta}{9}}\right)^2}\, .
\label{eq:deltaW}
\end{align}
In each case, a PT occurs if the nucleation condition $\Gamma(t_{\rm nuc}) = H^4(t_{\rm nuc})$ is satisfied, where the decay rate is faster than the background's expansion so that at least one bubble forms within a horizon for a constant Hubble rate, assuming that spacetime can be considered approximately static.

\subsection{Consistency analysis and phenomenological considerations} \label{sec:Consistency}
In addition to the PT signatures of this model, we are interested in its particle phenomenology. For a scalar with the potential (\ref{eq:Vphi}) to act as a spectator field during inflation and DM in the late universe, its couplings must obey certain conditions. The BSM potential energy should be subdominant to the inflaton's while $w(t)=-1$, $|V_{\phi}(\phi_{\rm inf}, h=0, R = 12 H_{\rm inf}^2)| < 3 M_P^2 H^2_{\rm inf}$, so that it does not affect the expansion dynamics. This leads to the following constraint for $\xi_{\phi}<0$,
\begin{align}
    \frac{\sigma^2}{\lambda^3_{\phi}} \left(1- \frac{4 C}{3} + \sqrt{1 - \frac{8 C}{9}} \right) \left( 1 + \sqrt{1 - \frac{8 C}{9}}\right)^{2} < 288 M_P^2 \left(\frac{H_{\rm inf}}{\sigma} \right)^2 \, ,
    \label{eq:spectator}
\end{align}
bounding the parameter space of the PT shown in Fig. \ref{fig:V-xi-negative}, whereas for $\xi_{\phi}>0$ the scalar lies at the origin , $V^{\xi_{\phi}>0}_{\phi}(\phi_{\rm inf}=0, h=0, 12 H_{\rm inf}^2)=0$. An upper bound can be placed on the portal coupling,
\begin{align}
    g < 0.13 \, ,
    \label{eq:g_BR}
\end{align}
from the Higgs branching ratio into invisible particles, if the BSM scalar is the product of the process \cite{Cosme:2018wfh}, and radiative corrections to its self-interaction must be subdominant to avoid fine-tuning,
\begin{align}
    \Delta \lambda_{\phi} \sim \frac{g^4}{16 \pi^2} < \lambda_{\phi} \, .
    \label{eq:Delta_lambda}
\end{align}

After kination, the BSM scalar settles at its high-field vev $\phi_{\rm reh}$ for both cases of positive and negative curvature couplings (bottom right panels of Figs. \ref{fig:V-xi-positive} and \ref{fig:V-xi-negative}). As time progresses and EWSB takes place, the Higgs field acquires its non-zero vev, $v \approx 246$ GeV, and the scalar acquires its mass via its Higgs portal, $m_{\phi}= g v /\sqrt{2}$.  Hence, the BSM scalar acts as DM in the late universe, if it matches its present day abundance $\Omega_{\phi, 0} \simeq 0.26$ \cite{Husdal:2016haj}, meaning that the portal coupling obeys
\begin{align}
    g = \frac{\sqrt{12 \Omega_{\phi, 0}} H_0 M_P}{v} \left( \frac{g_{* \mathrm{reh}}}{g_{*0}}\right)^{\frac{1}{2}} \left( \frac{T_{\mathrm{reh}}}{T_0}\right)^{\frac{3}{2}} \phi_{\rm reh}^{-1} \simeq 3 \times 10^{-17} \left(\frac{T_{\mathrm{reh}}}{80 \, \mathrm{GeV}}\right)^{\frac{3}{2}}  \frac{\lambda_{\phi}}{\sigma} \,.
    \label{eq:g_DM}
\end{align}
Considering the viable BSM parameter space for strong GW signals (see next section), $\sigma \gtrsim H_{\rm inf}$ and $\lambda_{\phi} \lesssim 10^{-4}$, Eq. (\ref{eq:g_DM}) suggests that the Higgs-portal is extremely weak. This implies that Eqns. (\ref{eq:g_BR}) and (\ref{eq:Delta_lambda}) are always satisfied, but moreover, that curvature-induced EWSB can occur for sufficiently low (re)heating temperatures, $T_{\rm reh} \leq 80$ GeV. With the decreasing magnitude of $R$ after inflation, the contribution from the non-minimally coupled Higgs term $\xi_h R/ |\lambda_h|$ to $v^2$ becomes subdominant and the EW symmetry breaks, if the dark scalar's contribution, $(g \phi_{\rm reh})^2/2|\lambda_h|$, is suppressed. The resulting bound always complies with Eq. (\ref{eq:g_DM}), for the highest value of the Higgs self-interaction $\lambda_h(\mu_{\rm EW})\approx 0.13$ \cite{Mantziris:2020rzh}, with a slightly stricter bound for $\lambda_h(\mu_{\rm inf} \gg \mu_{\rm EW})$,
\begin{align}
    g < \sqrt{\frac{2 |\lambda_h| \lambda_{\phi}}{\sigma}} v < 125 \sqrt{\frac{\lambda_{\phi}}{\sigma}} \, .
\end{align}

\section{Gravitational waves from bubble collisions in vacuum} \label{sec:GWs}

\begin{figure}[ht!]
\centering
\includegraphics[width=0.92\linewidth]{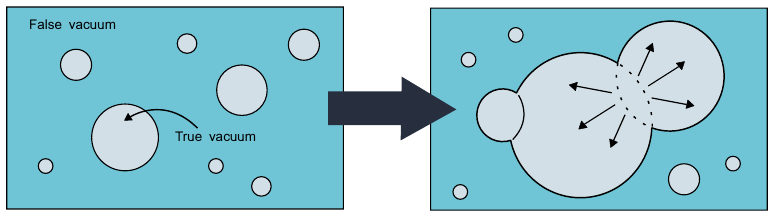}
\caption{Schematic depiction of gravitational wave emission from bubble collisions after a first-order phase transition of a scalar field. Credit: D. Weir \cite{Weir:2019thg}.}
\label{fig:Bubbles}
\end{figure}

After a bubble forms and expands in a false vacuum universe, it collides with other nucleated bubbles, as shown in Fig. \ref{fig:Bubbles}. When the bubbles fill the Hubble patch, the dark scalar lies at its high-field vev within the horizon. During this period of bubble percolation, the PT's latent heat, i.e. the energy released by the PT compared to the background's,
\begin{align}
    \alpha \equiv \frac{\rho_{\rm PT}}{\rho_{\rm kin}} = \frac{|\Delta V|}{3 M_P^2 H_{\rm inf}^2} \, ,
\end{align} 
is transferred to GWs via collisions, since there is no intermediate thermal plasma to facilitate GW generation from friction or turbulence \cite{Lewicki:2023ioy}. The strength of strong first-order PTs typically lies in the range $10^{-4}\lesssim \alpha \lesssim 10^{-1}$, where smaller values resemble second-order transitions and higher orrespond to supercooled PTs, an effect that complicates the analysis and was beyond the scope of our study \cite{Athron:2022mmm}. For the PT to proceed faster than the spacetime's expansion so that the bubbles can successfully percolate, a hierarchy between the timescales of each process emerges, $\beta_{\rm col} > \beta_w$. More specifically, the inverse time interval between bubble nucleation and percolation,
\begin{align}
    \beta_{\rm col}^{\xi_{\phi}<0} &= \eval{\sqrt{ \frac{d^2 S_4^E}{dt^2}}}_{t=t_{\rm col}} \, ,\\
    \beta_{\rm col}^{\xi_{\phi}>0} &= - \eval{ \dv{t} S_4^E }_{t=t_{\rm col}} \, ,
\end{align}
should be greater than the transition from inflation to kination, which depends on the inflationary model and was set to $\beta_w= 10 H_{\rm inf}$ for simplicity (see Fig. \ref{fig:inf}). \\

The gravitational ripples travel unobstructed after emission, carrying the signatures of their violent origin in both their amplitude and peak frequency , that redshift according to
\begin{align}
    \Omega_{\rm GW, 0} (f) &= \frac{1.67 \times 10^{-5}}{h^2} \left(\frac{H_{\rm col}}{H_{\rm reh}}\right)^{\frac{2\left(3w_{\rm int}-1\right)}{3w_{\rm int}+3}}  \left(\frac{H_{\rm col}}{\beta_{\rm col}}\right)^{2} \left(\frac{\alpha}{\alpha+1}\right)^2 \mathcal{S}(f) \, , 
    \label{eq:OmegaGW}\\
    f_{\rm peak, 0} &= 1.65 \times 10^{-5} \left(\frac{H_{\rm col}}{H_{\rm reh}}\right)^{\frac{3w_{\rm int}-1}{3w_{\rm int}+3}} {\left(\frac{ f_{\rm peak, col}}{H_{\rm col}}\right)} \left(\frac{T_{\rm reh}}{100 \mbox{ GeV}}\right) \, ,
\end{align}
where $ w_{\rm int} \approx 1$ is the time integrated EoS parameter with extended kination, $f_{\rm peak, col}=0.13 \beta_{\rm col}$, and $t_{\rm nuc} \approx t_{\rm col}$ assumes that bubble percolation takes place immediately after nucleation in cosmological time \cite{Caprini:2018mtu, Allahverdi:2020bys}. The fraction of the energy density between GWs and the rest of the universe (\ref{eq:OmegaGW}) follows a broken power-law spectral function, determined by numerical simulations \cite{Lewicki:2019gmv, Lewicki:2020azd},
\begin{align}
    \mathcal{S}\qty(f) = 25.09  \qty[ 2.41 \qty(\frac{f}{ f_{\textrm{peak},0} })^{-0.56} + 2.34 \qty(\frac{f}{ f_{\textrm{peak},0} })^{0.57} ]^{-4.2} \, .
\end{align}
A thorough analysis of the transition from the post-inflationary universe to the radiation domination epoch went beyond our scope, and we therefore made the simple assumption of instantaneous thermalisation, $H_{\rm reh}  = \frac{\pi}{3} \sqrt{\frac{g_{*\mathrm{reh}}}{10}} \frac{T_{\rm reh}^2}{M_{P}} \approx 9 \times 10^{-15}$ GeV for $T_{\rm reh}=80$ GeV, where $ g_{* \mathrm{reh}} = 106.75$ are the effective degrees of freedom, with the inflaton decaying to SM particles before the HBB.

\begin{figure}[ht!]
\begin{minipage}{.5\linewidth}
\centering
\subfloat{\includegraphics[scale=.72]{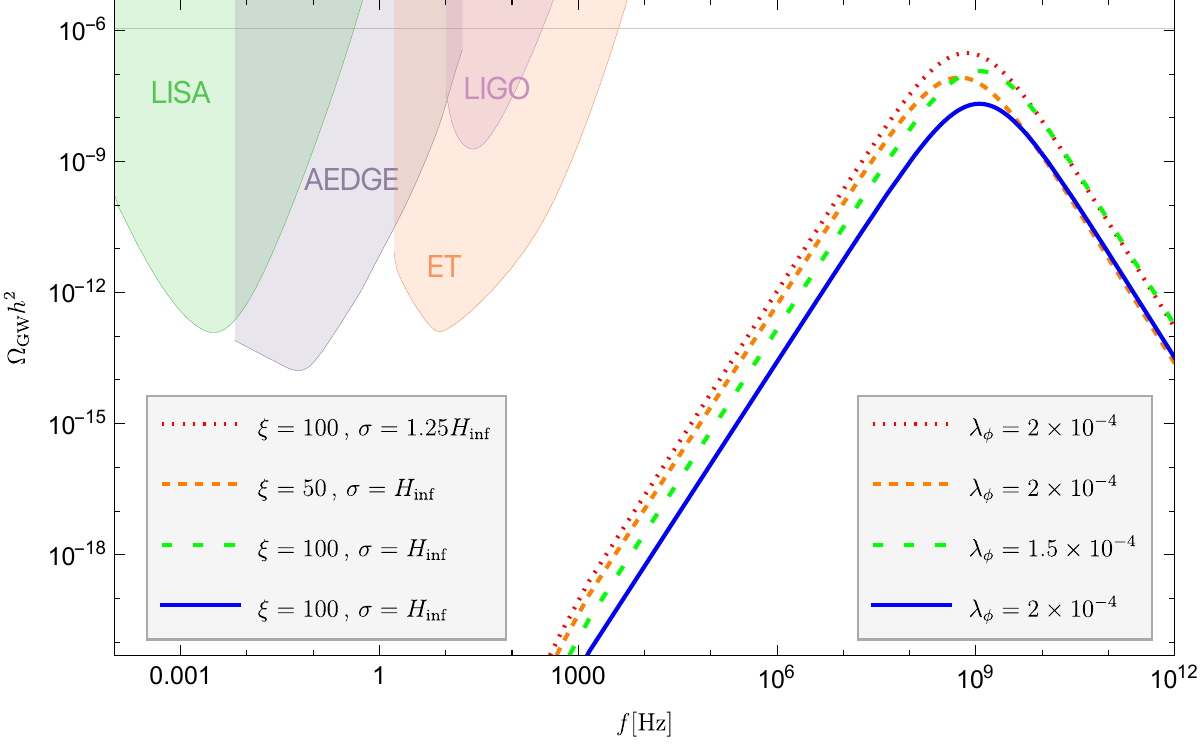}}
\end{minipage} \par\medskip
\begin{minipage}{.5\linewidth}
\centering
\subfloat{\includegraphics[scale=.72]{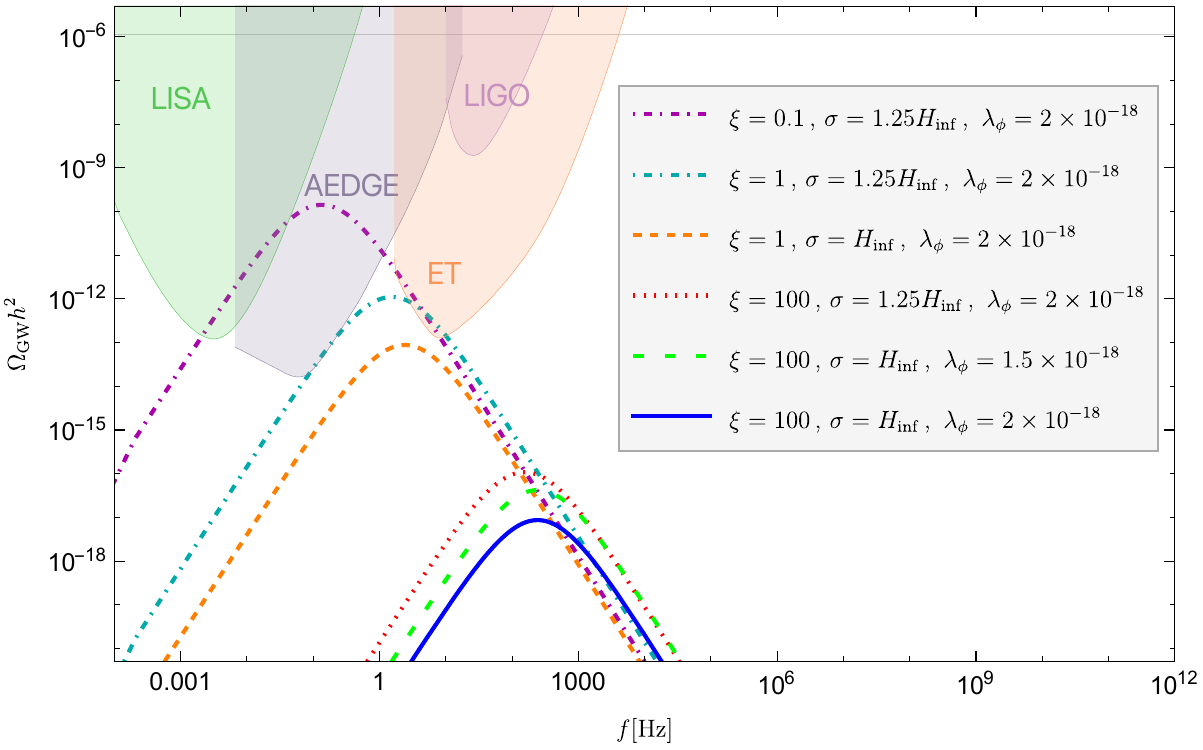}}
\end{minipage}
\caption{Gravitational wave spectra from phase transitions for $\xi_{\phi}>0$ with high scale inflation $H_{\rm inf}=10^{12}$ GeV (top) or low scale inflation $H_{\rm inf}=10^{-8}$ GeV (bottom), and projected detector sensitivities.}
\label{fig:GW_xi_positive}
\end{figure}

\begin{figure}[ht!]
\begin{minipage}{.5\linewidth}
\centering
\subfloat{\includegraphics[scale=.72]{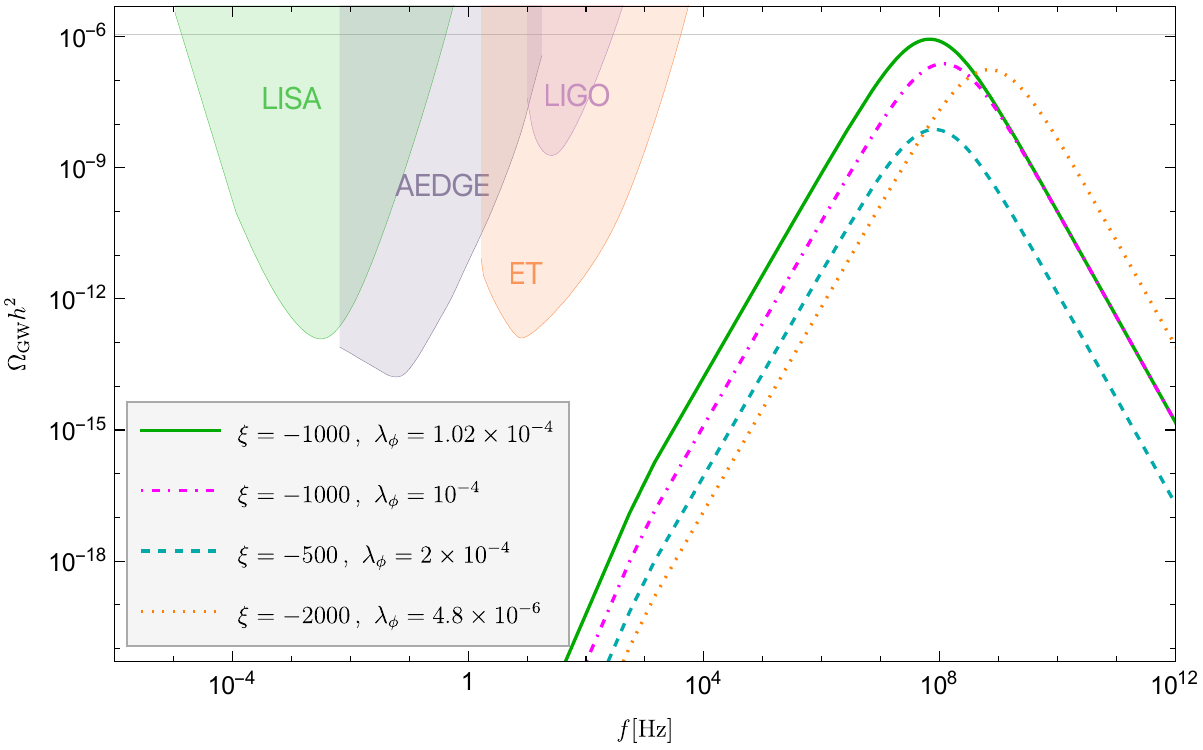}}
\end{minipage} \par\medskip
\begin{minipage}{.5\linewidth}
\centering
\subfloat{\includegraphics[scale=.72]{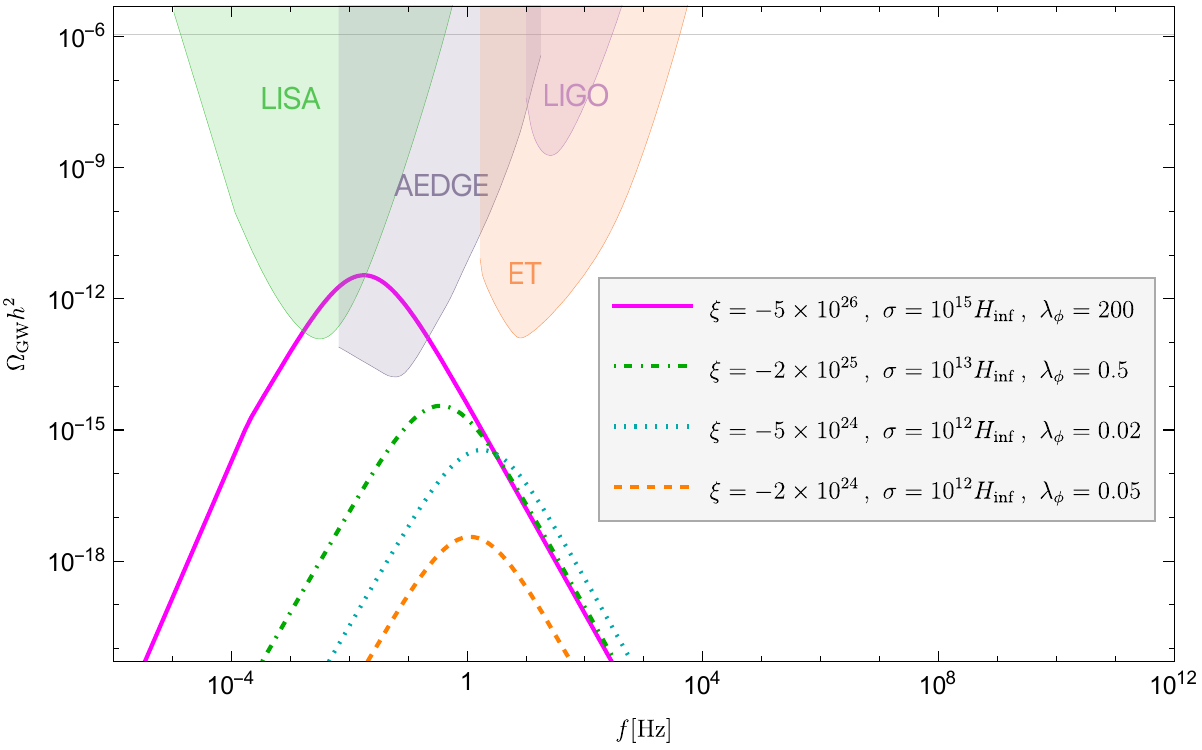}}
\end{minipage}
\caption{Gravitational wave spectra from phase transitions for $\xi_{\phi}<0$ with high scale inflation $H_{\rm inf}=10^{12}$ GeV and $\sigma = H_{\rm inf}$ (top) or low scale inflation $H_{\rm inf}=10^{-8}$ GeV (bottom), and projected detector sensitivities.}
\label{fig:GW_xi_negative}
\end{figure}

We showcase the resulting GW spectra for the favourable areas of the BSM parameter space in Figs. \ref{fig:GW_xi_positive} and \ref{fig:GW_xi_negative}. Starting with $\xi_{\phi}>0$ at the high inflationary scales of ``new'' and quintessential-like inflationary models, the signals lie at high frequencies with boosted amplitudes that approach the Big Bang Nucleosyntheis (BBN) bound at $\Omega_{\rm BBN} h^2 = 1.12 \times 10^{-6}$ (top panel of Fig. \ref{fig:GW_xi_positive}). The lowest inflationary scale above the EW pushes the spectra to lower frequencies while decreasing their amplitude and requiring extremely weak BSM self-interactions to adhere to all constraints and maintain high amplitudes (lower plot of Fig. \ref{fig:GW_xi_positive}). Similarly for $\xi_{\phi}<0$, strong GHz signals stem from high inflation scales (top panel of Fig. \ref{fig:GW_xi_negative}), whereas the spectra lie at lower frequencies for very low inflation scales (bottom plot of Fig. \ref{fig:GW_xi_negative}). In the later case, such inflationary scales are difficult to motivate theoretically and achieve from realistic model-building, while also the corresponding BSM curvature and cubic couplings are extremely strong, in order to comply with all the constraints and provide the strongest possible signatures. Finally, note that all spectra have a small tilt in their low-frequency tail from extended kination affecting the dilution of their energy density \cite{Gouttenoire:2021jhk}.

\section{Summary and future prospects}

This work serves as an overview of Ref. \cite{Mantziris:2024uzz}, where the prospect for GWs from a curvature-induced PT of a Higgs-portal DM scalar was studied. Our aim was to provide a GW signature to the model from Ref. \cite{Cosme:2018wfh}, via the non-thermal PT mechanism of Ref. \cite{Kierkla:2023uzo}, to preserve the theory's capability of hosting a non-thermal EWSB. The parameter space complying with all the conditions for a strong first-order PT, sufficient DM abundance, and curvature-induced EWSB assuming $T_{\rm reh}\leq 80$ GeV reads
\begin{align}
    10^{-32} \lesssim g \lesssim 10^{-21} \, , \qquad -10^3 \lesssim \xi_{\phi} \lesssim 10^2 \, , \qquad \sigma \gtrsim H_{\rm inf} \, , \qquad \lambda_{\phi} \lesssim 10^{-4} \, ,
\end{align}
resulting in an extremely small mass for the dark scalar,  $ 10^{-30} \textrm{ GeV} \, \lesssim m_{\phi} (T_{\rm reh}) \lesssim 10^{-19} \textrm{ GeV}$. PTs at the typical $10^8-10^{12}$ GeV inflationary scales lead to GWs lying beyond current and upcoming detector sensitivities, in contrast to PTs where inflation takes place close to the EW scale. However, strong GHz signals do not suffer from astrophysical and galactic interference and can only be attributed to a primordial source \cite{LISACosmologyWorkingGroup:2022jok}. These symmetrical spectra from cold PTs can be easily differentiated from their thermal counterparts with the pronounced high-frequency tails due to the plasma interactions \cite{Cembranos:2024pvy}. Finally, extended kination imprints a ``smoking-gun'' characteristic in spectra produced in the very early universe, in the form of a low-frequency tilt.

Although the detection prospects for GWs of this theory fall on future high frequency experiments, see for example \cite{Aggarwal:2020olq, Bringmann:2023gba, Blas:2026ybh} and references therein, it is exciting that such a generic PT mechanism can be applied to various BSM frameworks in the early universe and provide insights into the properties of quantum fields and the primordial expansion history. However, there is no direct correlation between GWs signals and the DM model, besides constraining the relevant parameter space of the couplings, and thus, this mechanism can only complement and not guide DM model building. In any case, we believe that there is a promising and multifaceted outlook for future studies regarding different BSM and inflationary models, with the prospect of multiple PTs from the oscillating Ricci scalar being particularly interesting. Furthermore, it is essential to properly incorporate the transition from kination to (re)heating for precision studies in future investigations \cite{Barman:2026kab}. Finally, we anticipate that expanding this formalism to non-minimally coupled matter-curvature theories would offer novel insights into their dynamics, considering the central role that spacetime curvature plays in these frameworks \cite{March:2025uqt, BarrosoVarela:2025yyr, Lobo:2025eex, Varela:2024egg, Carloni:2024ybx, Bertolami:2017svl}.

\break

\section*{Acknowledgements}
The author acknowledges the collaboration of Orfeu Bertolami and the support of FCT - Fundação para a Ciência e Tecnologia, I.P. through the project CERN/FIS-PAR/0027/2021, with DOI identifier 10.54499/CERN/FIS-PAR/0027/2021, that resulted in Ref. \cite{Mantziris:2024uzz}. This study was presented at Corfu Summer Institute 2025 with the support from the FCT project with DOI identifier 10.54499/2024.00252.CERN. The author is grateful to the Theoretical Physics group of the University of Ioannina for the hospitality during his visit, when this manuscript was completed.

\end{document}